\documentclass{article}

\usepackage{arxiv}

\usepackage{amsmath}

\usepackage[utf8]{inputenc} 
\usepackage[T1]{fontenc}    
\usepackage{hyperref}       
\usepackage{url}            
\usepackage{booktabs}       
\usepackage{amsfonts}       
\usepackage{nicefrac}       
\usepackage{microtype}      
\usepackage{cleveref}       
\usepackage{lipsum}         
\usepackage{graphicx}
\usepackage[square,numbers,super,sort&compress]{natbib}
\usepackage{doi}

\newcommand{\FIGURESDIR}{.}

\newcommand{\adpaperfigz}{\FIGURESDIR}

\newcommand{\ADBASEMODEL}{PDCCore}
\newcommand{\ADHGMODULE}{PopuSense}
\newcommand{\RESNET}{ResNet101}

\newcommand{\CGVANILLA}{\ADBASEMODEL}
\newcommand{\CGLOWPOP}{Narrow \ADHGMODULE}
\newcommand{\CGHIGHPOP}{Wide \ADHGMODULE}

\newcommand{\CONTRASTBASED}{contrast}
\newcommand{\TEXTUREBASED}{texture}

\newcommand{\NMRINORMAL}{481\space}
\newcommand{\NNORMAL}{3515\space}

\newcommand*{\zsubsection}{\subsection}

\usepackage{booktabs, threeparttable} 
\usepackage{tabularx}
\usepackage{caption}

\crefname{section}{Sec.}{Secs.}
\Crefname{section}{Section}{Sections}
\Crefname{table}{Table}{Tables}
\crefname{table}{Tab.}{Tabs.}
\newcommand{\ztitle}{Harnessing Intra-group Variations Via a Population-Level Context for Pathology Detection} 
\newcommand{\zruntitle}{\ztitle}

\title{\ztitle}

\date{}

\usepackage{authblk}

\newbox{\orcid}\sbox{\orcid}{\includegraphics[scale=0.06]{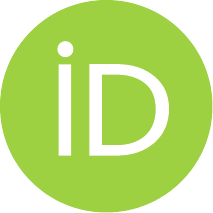}} 
\author[1]{%
	\hspace{1mm}P.~Bilha Githinji
} 
\author[1]{\hspace{1mm}Xi Yuan}
\author[1]{\hspace{1mm}Zhenglin Chen }
\author[1]{\hspace{1mm}Ijaz Gul }
\author[1]{\hspace{1mm}Dingqi Shang }
\author[1]{\hspace{1mm}Wen Liang }
\author[1]{\hspace{1mm}Jianming Deng }
\author[2]{\hspace{1mm}Dan Zeng }
\author[3]{\hspace{1mm}Dongmei Yu }
\author[4]{\hspace{1mm}Chenggang Yan}
\author[1]{%
	\hspace{1mm}Peiwu Qin
}
\affil[1]{Tsinghua University, Tsinghua-Berkeley Shenzhen Institute, China}
\affil[2]{School of Communication and Information Engineering, Shanghai University, China}
\affil[3]{School of Mechanical, Electrical \& Information Engineering, Shandong University, China}
\affil[4]{School of Automation, Hangzhou Dianzi University, China}

\renewcommand{\headeright}{} 
\renewcommand{\undertitle}{} 
\renewcommand{\shorttitle}{\zruntitle}

\hypersetup{
pdftitle={\ztitle},
pdfsubject={},
pdfauthor={P.~Bilha Githini, et al.},
pdfkeywords={Pathology Detection, Anomaly Detection, Latent Code Refinement, Population-level Context, Hypergraph},
}

\begin{document}

\maketitle

\begin{abstract}	

Realizing sufficient separability between the distributions of healthy and pathological samples is a critical obstacle for pathology detection convolutional models. Moreover, these models exhibit a bias for contrast-based images, with diminished performance on texture-based medical images. This study introduces the notion of a population-level context for pathology detection and employs a graph theoretic approach to model and incorporate it into the latent code of an autoencoder via a refinement module we term PopuSense. PopuSense seeks to capture additional intra-group variations inherent in biomedical data that a local or global context of the convolutional model might miss or smooth out. Proof-of-concept experiments on contrast-based and texture-based images, with minimal adaptation, encounter the existing preference for intensity-based input. Nevertheless, PopuSense demonstrates improved separability in contrast-based images, presenting an additional avenue for refining representations learned by a model.
		
\end{abstract}


\section{Introduction}
\label{sec:adpintro}

The dichotomy between \CONTRASTBASED-based and \TEXTUREBASED-based images is inadvertently apparent in anomaly or pathology detection models employing auto-encoders, variational encoders and generative adversarial models, which we collectively refer to as pathology detection convolutional models (PDCs). Evaluation of these models on medical images highlights a notable preference for intensity-driven variations, as evidenced by the relatively superior performance of state-of-the-art (SOTA) models on \CONTRASTBASED-based images~\cite{tschuchnig_anomaly_2022,bercea_generalizing_2023,lagogiannis_unsupervised_2024}.

Moreover, empirical evidence shows that PDCs may still reconstruct anomalous regions, hindering the separability of healthy and pathological distributions, and resulting in high false positive rates~\cite{tschuchnig_anomaly_2022,bercea_generalizing_2023,bercea_what_2023,lagogiannis_unsupervised_2024}. One of the cited challenges lies in striking a balance between learning a compact latent code and one that is also sufficiently informative to prevent the reconstruction of out-of-distribution data points. Relatedly, medical images exhibit internal variations within subjects and lesion types that are typically smoothed out in favor of population-level factors. As a result, a compact and informative code for a PDC would not only go beyond intensity-driven optimization but also incorporate the internal diversity within a healthy population. 

This study models a population-level context for pathology detection, employing a hypergraph in the latent space of a PDC to harness underutilized associations and augment the latent code of the PDC.  We term the associated module PopuSense and apply it to both intensity-based and texture-based datasets to explore its utility under the different input types. Specifically, this research

\begin{itemize}
	\item{Introduces a latent code refinement module (PopuSense) that embodies a population-level context in the latent space of a pathology detection convolution autoencoder and models it using higher-order relational associations.}
	
	\item{Conducts experiments on the proposed method applying it to a brain tumor dataset and a retinal fundus dataset, which are intensity-based and texture-based inputs respectively.}
	
	\item{Under vanilla conditions, empirical results demonstrate improvement for intensity-based input and a challenge extending similar performance to textural input, a recurring theme in existing research. }
\end{itemize}

\section{Background}
\label{sec:adplit}

The core structure of a pathology detection convolutional model (PDC) is a standard auto-encoder convolution model. Variational autoencoders (VAEs) and generative adversarial networks (GANs) extend the autoecoder (AE) and are capable of image synthesis. These models have been utilized in pathology or anomaly detection to estimate the latent representation of normal or healthy distributions, and then determine anomalous or out-of-distribution observations based on the magnitude of the reconstruction error~\cite{bulusu_anomalous_2020,tschuchnig_anomaly_2022}. 

To realize a latent code that is both concise and sufficiently informative, some methods employ input-centric tactics to make the model resilient to spurious perturbations in the input~\cite{tan_self-supervised_2023,zhou_self_2023,fragemann_review_2022,tschuchnig_anomaly_2022}. Other methods incorporate auxiliary branches into the model, providing control, conditioning or guidance to facilitate learning a representation that aligns with some desired prior such as structural correspondence~\cite{zhou_memorizing_2022,zhou_proxy-bridged_2022,fragemann_review_2022,tschuchnig_anomaly_2022,liu_prompt-enhanced_2023}. Furthermore, some techniques entail learning a dictionary (also referred to as a codebook or memory bank) of latent codes making the PDC model adaptable to variations or nuances in the data~\cite{gong_memorizing_2019,ilie-ablachim_anomaly_2022,zhou_proxy-bridged_2022}. 

PDCs, by employing convolutional neural networks (CNNs), inherently entail a local spatial context by virtue of the convolution kernel and operation, and have been extended to capture global spatial contexts via techniques such as dilated kernels, skip connections, and attention modules~\cite{alzubaidi_review_2021}. By learning a set of possible latent codes, the dictionary-learning technique seems to factor in an additional context; a population- or sample-level context. We envisage this context as a potential avenue for accommodating intra-group heterogeneities in medical data. 

Nevertheless, learning a dictionary of latent codes increases model parameters as a function of the codebook size, complicating training and data needs~\cite{ilie-ablachim_anomaly_2022}. Additionally, determining an optimal codebook size and promoting its diversity may call for considerable hyper-parameter tuning and empirical analysis~\cite{ilie-ablachim_anomaly_2022}. Instead, we consider a graph theoretic approach to capture and incorporate the relational structure of hidden or underutilized associations in the population context.

Graph learning concepts offer adaptable frameworks for addressing graph data as well as modeling relationships in non-graphical domains~\cite{xia_graph_2021,ma_comprehensive_2021,dai_hypergraph_2023,ahmedt-aristizabal_survey_2022}. Moreover, they facilitate a mathematical framework for representing joint probability distributions when combined with probability theory~\cite{dietrichstein_anomaly_2022}. We leverage a hypergraph to incorporate relational assumptions in a non-graphical domain for the purposes of enhancing representations learned by a pathology detection model, fostering the separability of healthy and pathological reconstruction errors.

\section{Method}
\label{sec:adpmethod}

Motivated by the notion of a population-level context and its potential utility, we integrate it into a reasonably capable PDC. This section details the associated model design and evaluation approach. 
 
\subsection{Model}
\label{sec:adpmethod-model}

\begin{figure}
	\label{admodel}
	\centering
	\small 
	\includegraphics[width=0.85\textwidth]{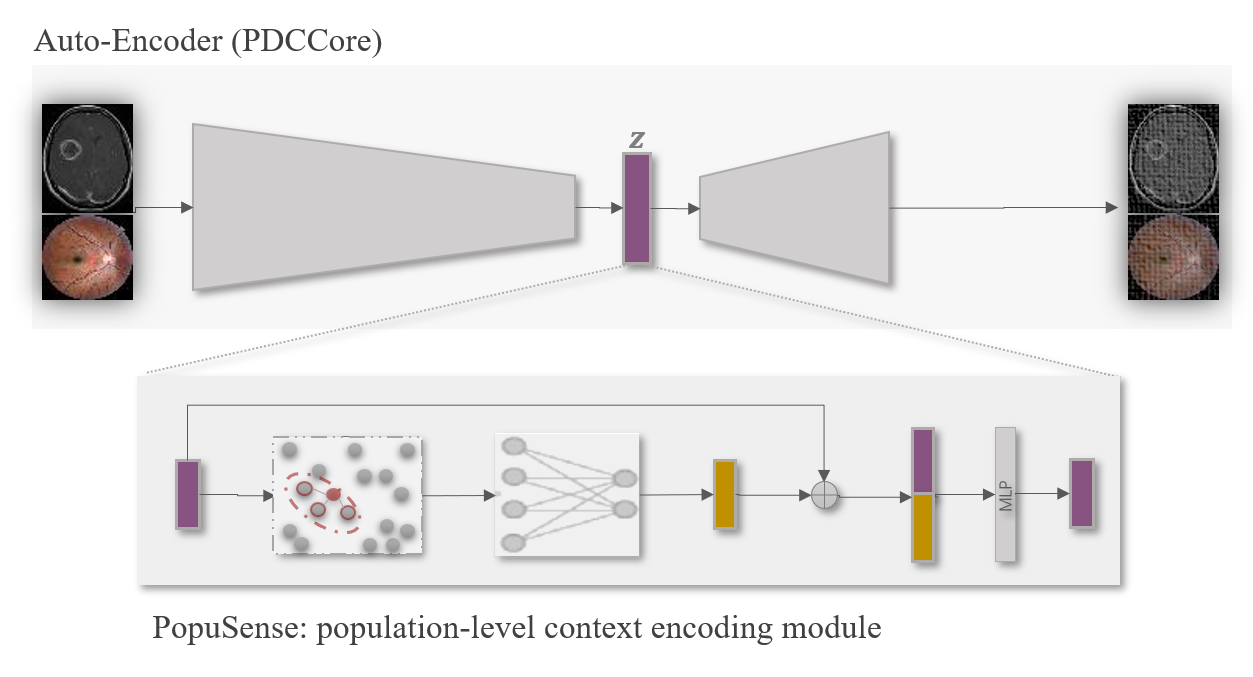}
	\caption{The latent code of a pathology detection autoecoder is augmented with a population-level context to leverage intricate or hidden associations within a group of similar observations. PopuSense, the corresponding module, takes as input the latent code of the autoencoder and estimates a hypergraph to capture higher-order relationships within a group. Afterward, a hypergraph convolutional neural network learns the associated embedding, which is then incorporated into the original latent code via a single fully connected layer.}
\end{figure}

The proposed method operates within the latent space of a pathology detection convolutional model (PDC). We refer to the core architecture of the PDC as the \ADBASEMODEL~and to the module that refines the latent code as the \ADHGMODULE~module. Figure \Cref{admodel} illustrates the model components. 

\textbf{PDCCore:} This part of the model entails a convolutional auto-encoder (AE) with a tunable bottleneck. A \RESNET~backbone is adopted for the encoder function of the AE and augmented with multi-scale features to enhance the detection of small lesions. A bottleneck module is appended to further constrain the size of the latent code. 

In contrast, the decoder function is underspecified relative to its partner encoder by having fewer layers and not incorporating skip connections from all the layers of the encoder. Specifically, the additional bottleneck component of the encoder does not have a matching module in the decoder, and instead, the decoder simply upsamples through bilinear interpolation. In addition, the decoder contains no skip connections from the early layers of the encoder, which capture lower-level features. 

Consequently, the AE has a powerful encoder for our medical images and the asymmetry regularizes against learning an identity function. Moreover, for additional regularization, a denoising layer is introduced just before the encoder and initially set to a dropout unit for simplicity (this is not a fixed component and can be adapted as desired).

\begin{figure}[h!]
	\centering
	\small 
	\includegraphics[width=.70\textwidth]{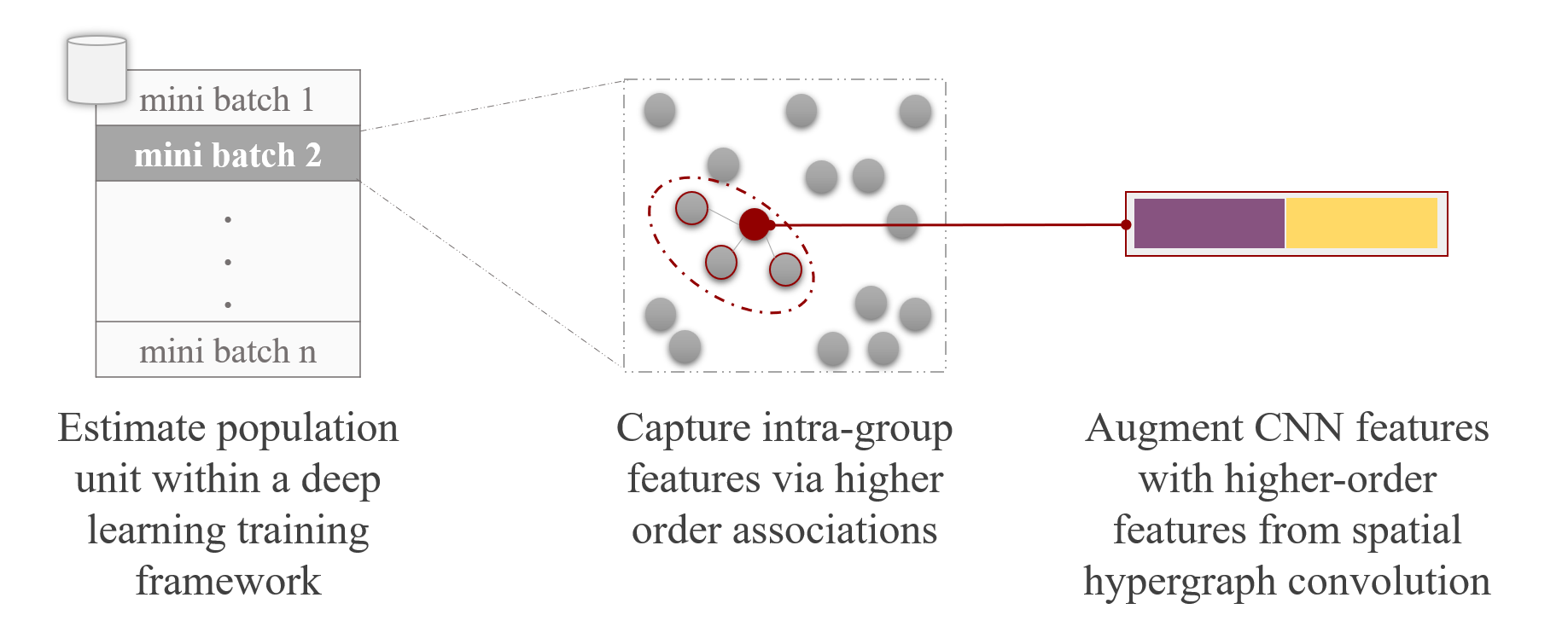}
	\caption{Overview of the population-sensitive latent code refinement approach.}
	\label{adoverview}
\end{figure}

\textbf{PopuSense:} This part of the model employs a graph-theoretic approach to represent and encode a population-level context for normal samples. 
To operate within a deep learning training framework, a population context is proxied by training mini-batches. Observations leverage higher-order associations among similar instances in a mini-batch and the PDC latent representation is accordingly augmented with the hypergraph embedding. This is illustrated in~\Cref{adoverview}.

Firstly, a population-level similarity context is modeled using a hypergraph. The observations in a batch are depicted as nodes while similarity scores between the latent codes from the \ADBASEMODEL~correspond to edges. Similarity scores are computed as the L2 norm between latent codes and edges are only activated for the top k scores. The result is a binary incidence matrix $H \in {R}^{b \times k}$ and an associated weight matrix $W = I_{k}$ that is equivalent to the identity matrix as our hypergraph is unweighted. The dimensions $b$ and $k$ correspond to the number of observations in a batch and the size of a neighborhood.

Thereafter, we apply a graph convolutional neural network to the latent codes of the base model, denoted $Z_{cnn}$, and the estimated hypergraph $H$. Finally, the initial latent codes are augmented with the output of the graph convolution using a single fully connected layer and projected to a lower dimension of the same size as the bottleneck of the base model \ADBASEMODEL, producing the enhanced latent code. \Cref{eq-popusense} depicts this process, where the component $D_{v}^{-\frac{1}{2}} H W D_{e}^{-1} H^T D_{v}^{-\frac{1}{2}}$ is derived from the hypergraph Laplacian $L = I - D_{v}^{-\frac{1}{2}} H W D_{e}^{-1} H^T D_{v}^{-\frac{1}{2}}$ and serves as a transition matrix for propagating messages during the graph convolution operation. In addition, $Z_{pop}$ is the embedding of the encoded population-level context, $HGC$ denotes the hypergraph convolution operation, $P$ is the projection matrix from layer $l$ to $l+1$, and $\sigma$ and $\sigma_{hgc}$ are nonlinearity functions.

\begin{equation}
	\label{eq-popusense}
	\begin{aligned} 
		Z^{l+1} &= MLP\big( CONCAT\big( Z_{cnn}^l, Z_{pop}^l\big)\big) \\
		&= \sigma \big( (Z_{cnn}^l || HGC(Z_{cnn}^l) ) P^{l} \big) \\
		&=\sigma \big( \big(Z_{cnn}^l || \sigma_{hgc} \big( Z_{cnn}^{l} \theta_{self}^{l} + D_{v}^{-\frac{1}{2}} H W D_{e}^{-1} H^T D_{v}^{-\frac{1}{2}} Z_{cnn}^{l} \theta_{hood}^{l} \big) \big) P^{l} \big)\\
	\end{aligned}
\end{equation}

\textbf{Loss function:} The loss function, as outlined in equation \Cref{adloss}, combines the mean squared error (MSE) and the structural similarity index metric (SSIM). The SSIM offers a more comprehensive assessment of image quality since it considers luminance, contrast and structure. We empirically find that supplementing it with the MSE loss yields reconstructions with contrast and luminance that closely resemble the original input. The contribution of the two metrics can be adjusted using hyperparameters $\lambda_1$ and $\lambda_2$. In the equation, $x$ is the original input and $\hat{x}$ is the reconstructed output. There is no additional regularization for the latent code as we first explore the capacity of \ADHGMODULE~unconfounded.

\begin{equation}
	\label{adloss} 
	L(x, \hat{x}) = \lambda_1 \space SSIM(x, \hat{x}) + \lambda_2 \space MSE(x, \hat{x})
\end{equation}

\textbf{Initial prototype.} We initially simplify the encoder and decoder functions of the pathology detection autoencoder as well as the input it operates on. This enables quick exploration since PDCs are large models.  
The autoencoder functions entail a convolutional block comprising of a convolutional layer, batch normalization and a nonlinear activation. The input resolution is reduced to $32 \times 32$ and the loss function is the mean squared error (MSE) $(x - \hat{x})^{2}$ between the original input $x$ and its reconstruction $\hat{x}$. Furthermore, since a reduction in resolution considerably deteriorates a color fundus image and PDCs are known to favor intensity-based input, we prioritize the brain radiology images. The prototype models are trained for about 30 epochs with 310 images.

We primarily implement using PyTorch~\cite{paszke2017automatic} and leverage existing packages for the foundational elements in well-established tasks. The PyTorch Segmentation Models package~\cite{Iakubovskii:2019} provides access to the \RESNET~backone through its model zoo. To compute SSIM for both grayscale and color images, we utilize MS-SSIM~\cite{mssim.pytorch}. Furthermore, Deep Hypergraphs (DHG)~\cite{gao2022hgnn} serves as a PyTorch-compatible graph learning module that extends existing graph neural networks to hypergraphs.

\subsection{Evaluation}
\label{sec:adpmethod-eval}

The difference between an original input $x$ and its reconstruction $\hat{x}$, denoted as the residual $(x - \hat{x})$, represents the erroneous pixels, pinpointing pathological regions. Examining the distribution of these residuals provides insights into the model's ability to distinguish between healthy and pathological samples. Distribution plots characterize the residuals, and the goal is minimal overlap between the healthy and pathological distributions. Additionally, a hypothesis test assesses the statistical difference between the distributions.

Furthermore, thresholding the residuals can categorize samples as healthy or pathological, facilitating the exploration of false positive rates via the accompanying ROC (receiver operating characteristic) curves. Thresholds are based on the statistics of the training sample residuals.

\section{Experiments}
\label{sec:adpexper}

\zsubsection{Input setup}

Leveraging publicly available sources, we assemble two datasets corresponding to contrast and textural input types. The \CONTRASTBASED-based dataset is a public brain tumor dataset that is a collection of three other datasets and has diversity in intensity levels~\cite{noauthor_brain_nodate}. We select axial plane images and drop low-quality images that have been overly preprocessed, resulting in \NMRINORMAL images that are marked healthy. 
For the \TEXTUREBASED-based dataset, we adopt a retinal fundus public dataset comprising of a diverse collection of other datasets and with a unified definition of what is normal or healthy~\cite{githinji_irfundusset:_2024}. We access a total of \NNORMAL healthy images.

Images undergo noise reduction preprocessing using median blur and CLAHE, and get resized and center cropped to yield a final resolution of 160x160. In addition, the green color channel is selected for the fundus dataset and all images are standardized to the range [0,1]. Furthermore, sample sizes are bolstered via six offline augmentations that entail combinations of blurring, vertical and horizontal flipping, and random masking using a random number and size of square-shaped masks that are filled with the median value of the input image. 

The training phase exclusively utilizes healthy samples, while a combination of healthy and pathological images is used for evaluation. The healthy images are randomly split into 80\% training set and a 20\% test set.

\zsubsection{Comparison groups} 
As described in the method section, we build a reasonably capable autoencoder (\ADBASEMODEL) and then explore the contribution of \ADHGMODULE, which incorporates a population-based context. Three models are configured as per below and trained with a batch size of 32. 
\begin{itemize}
	\item {\textit{\CGVANILLA}: This represents a conventional pathology detection autoencoder and is the baseline case}
	
	\item {\textit{\CGLOWPOP}: Here we implement \ADHGMODULE~using 35\% of the batch-size as the number of closest neighbors to consider when representing the population-level context.}	
	
	\item {\textit{\CGHIGHPOP}: In this case \ADHGMODULE~uses 70\% of the batch-size as the number of closest neighbors to consider when representing the population-level context.}	
\end{itemize}

\subsection{Results and discussion}
\label{sec:adpdiscuss}

\begin{figure}[ht!]
	\centering
	\includegraphics[width=1.\textwidth]{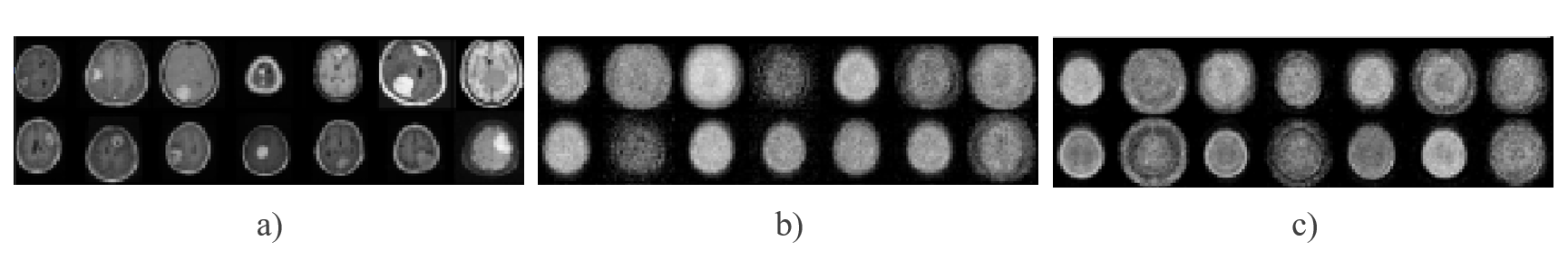}
	\caption{Prototype output. Figure a) is the original input, b) is the reconstruction by the vanilla autoencoder, and c) is the output when PopuSense is utilized. Additional detailing is observed in the outputs in c). Note that the reconstructions are for the normative distribution; anomalous regions should not be reconstructed.} 
	\label{fig:adtoy}
\end{figure}

\textbf{Initial prototype.}~\Cref{fig:adtoy} showcases sample outputs for a vanilla autoencoder model and the corresponding upgraded model that contains a PopuSense module. Qualitative inspection signals the averaging of features by the vanilla model and the stacking of information from observations within a subgroup by the upgraded model.

\begin{figure}[ht!]
	\centering
	\small 
	\includegraphics[width=0.95\textwidth]{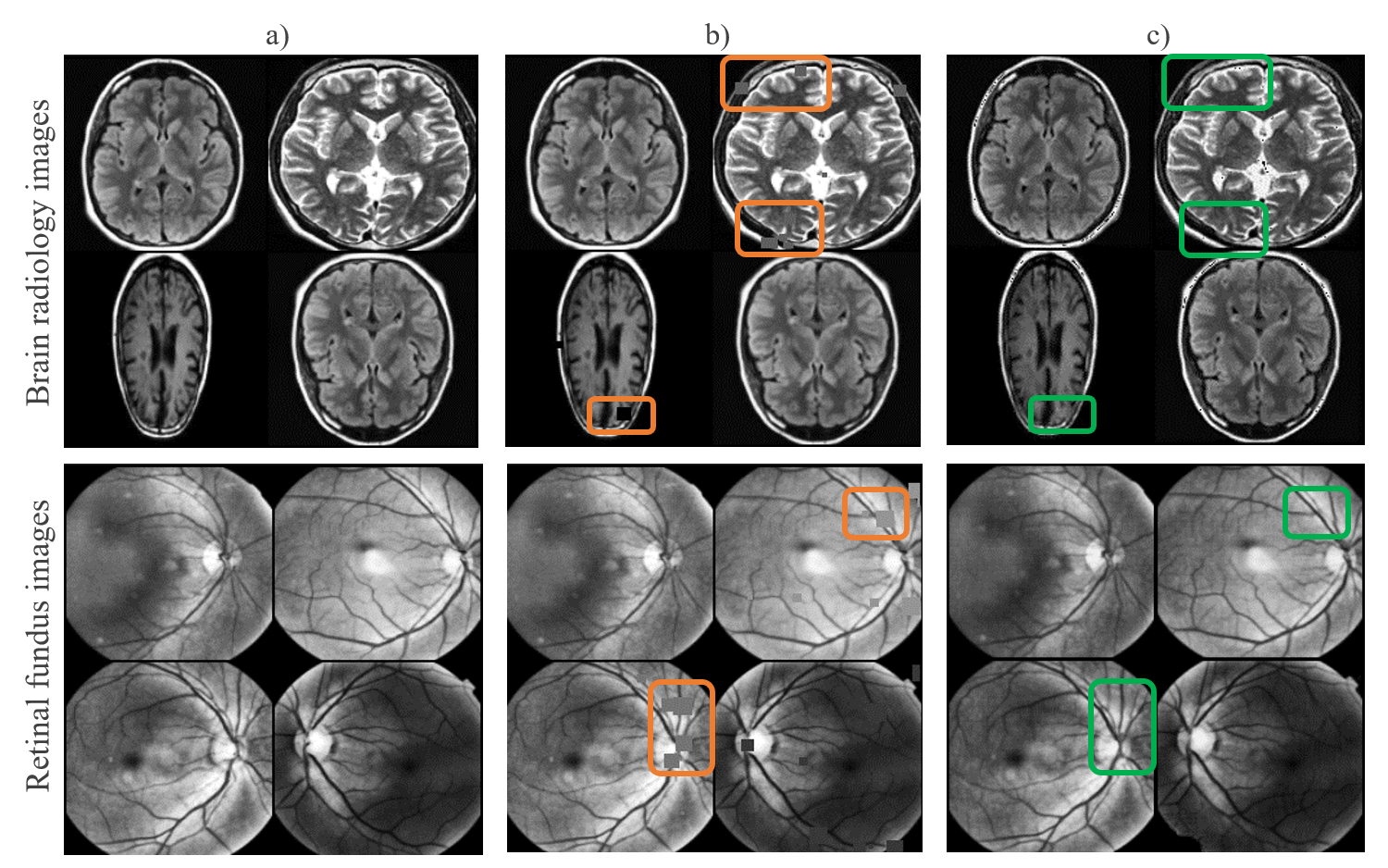}
	\caption{PDCCore capacity. Column a) corresponds to the original input, b) the augmented input, and c) the reconstructed output. The orange boxes highlight areas with masked augmentation, which are successfully reconstructed in the output images in c).}
\label{adcore-aug}
\end{figure}

\textbf{Capable PDCCore.} ~\Cref{adcore-aug} demonstrates the robustness of the PDCCore. Despite receiving augmented versions of the original images, the autoencoder successfully reconstructs the original input as qualitatively demonstrated by the completion of masked regions. 

\begin{figure}[ht!]
	\centering
	\small 
	\includegraphics[width=0.95\textwidth]{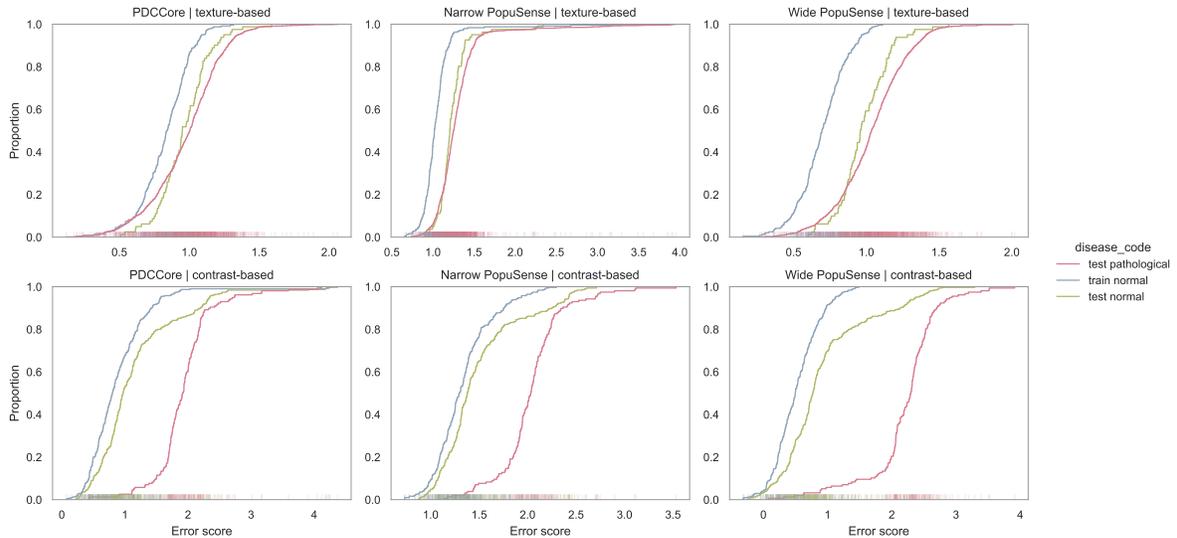}
	\caption{Empirical cumulative distribution plots for the error scores, showing the extent of overlap between healthy and pathological distributions for contrast and textural input, and across the three model configurations. Along the x-axis are the corresponding rug plots.}
	\label{adrez-distz-xe}
\end{figure}

\textbf{Separability of distributions.}~\Cref{adrez-distz-xe} presents distribution plots of the residuals or error scores for the two types of input and across the three model configurations. Ideally, we expect the distributions for new normal observations and training normal samples to align, while their overlap with the distribution of pathological samples is minimized. This expectation is evident for the \CONTRASTBASED-based dataset but not for the \TEXTUREBASED-based images. Additionally, the \CGHIGHPOP~configuration has the least overlap between healthy and pathological distributions for \CONTRASTBASED-based input, while \CGLOWPOP~has a distribution overlap pattern close to the baseline case (PDCCore). 

\begin{figure}[ht!]
	\centering
	\small 
	\includegraphics[width=0.85\textwidth]{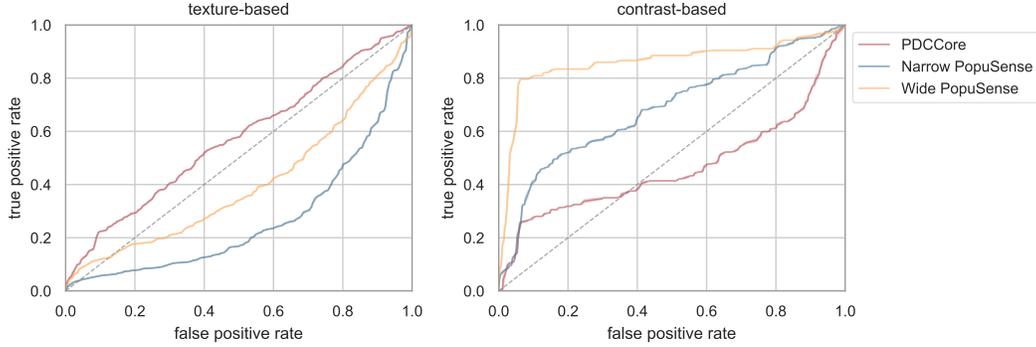}
	\caption{ROC curves from classifying samples as pathological for the three model configurations and two input types. }
	\label{adrez-thresh-roc}
\end{figure}

\Cref{adrez-thresh-roc} illustrates the ROC curves associated with classifying observations as healthy or pathological based on the thresholded error scores. The ideal model is characterized by points near the top-left corner, representing a good balance of high specificity (true positive rates) and reduced false positive rates. Similar to the results from the distribution plots, the \CONTRASTBASED-based dataset has relatively better performance under the \CGHIGHPOP~configuration. Performance under \TEXTUREBASED-based input appears no better than chance. 

\begin{figure}[ht!]
	\centering
	\small 
	\includegraphics[width=1.0\textwidth]{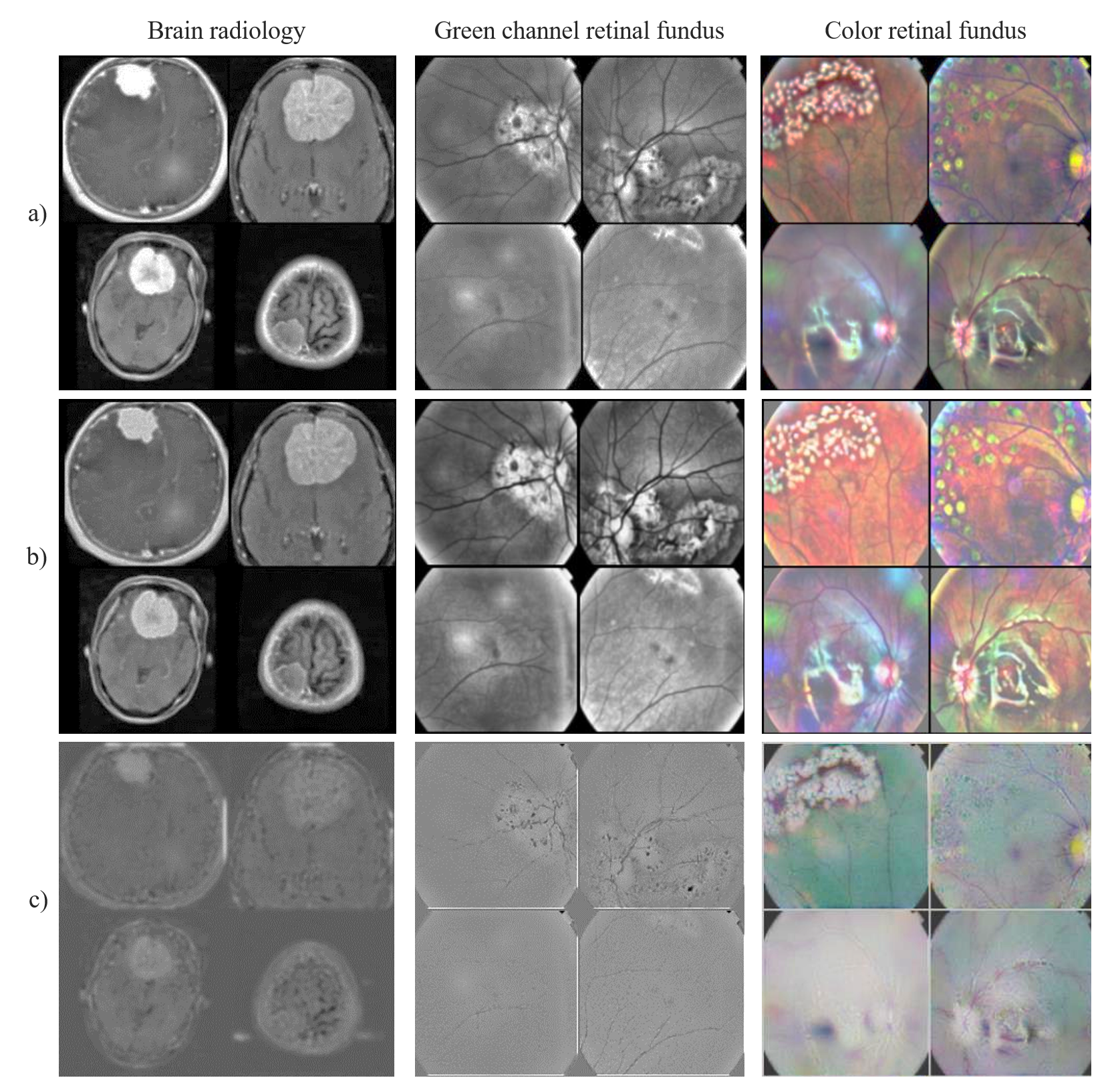}
	\caption{Reconstructed output for \CGHIGHPOP~configuration. Row a) shows the original images, b) the reconstructed output and c) the anomaly residuals or errors. Anomalous regions continue to be reconstructed and there seems to be an intensity-based differential in the residual maps in c)}
\label{adrez-reconz}
\end{figure}

Relatedly, the hypothesis tests reveal similar patterns. While most of the paired distributions are statistically different, contrast-based input under~\CGHIGHPOP~configuration has the largest statistically significant absolute difference between the mean residual error of the training samples and the mean residual error of the pathological samples, with a value of $8.654$ compared to $4.906$ for the configuration in second position. Moreover, a qualitative review of the reconstructed images reveals ongoing reconstruction of anomalous regions by the pathology detection convolution model. The hypothesis test results are tabulated in~\Cref{tbl:ad-kruskall}, while~\Cref{adrez-reconz} demonstrates select reconstructions.

\begin{table}[h!]	
	\centering
	\small
	\caption{\label{tbl:ad-kruskall} Hypothesis test on distribution pairs}
	\small\addtolength{\tabcolsep}{-3.05pt}
	\begin{threeparttable}[c]
		\begin{tabular}{llcccccccc}
\toprule
model & dataset & groups & $\mu_{1}$ & $\mu_{2}$ &  $|\mu_{1} - \mu_{2}|$ & $n_1$ & $n_2$ & t-stat & p-val \\
\midrule
Narrow PopuSense & Texture-based  & \footnotesize{TH Vs VP}  & 3.067 & 4.038 & 0.971 & 243 & 1066 & -3.720 & 0.000 \\
 & Texture-based  & \footnotesize{TH Vs VH} & 3.067 & 3.567 & 0.500 & 243 & 81 & -1.482 & 0.139 \\
 & Contrast-based  & \footnotesize{TH Vs VP}  & 3.985 & 8.182 & 4.198 & 310 & 157 & -18.366 & 0.000 \\
 & Contrast-based  & \footnotesize{TH Vs VH} & 3.985 & 4.910 & 0.925 & 310 & 310 & -5.649 & 0.000 \\
Wide PopuSense & Texture-based & \footnotesize{TH Vs VP}  & 2.039 & 2.901 & 0.862 & 243 & 1066 & -18.022 & 0.000 \\
 & Texture-based & \footnotesize{TH Vs VH} & 2.039 & 2.717 & 0.678 & 243 & 81 & -13.439 & 0.000 \\
 & Contrast-based & \footnotesize{TH Vs VP}  & 1.790 & 10.443 & \textbf{8.654} & 310 & 157 & -25.557 & 0.000 \\
 & Contrast-based & \footnotesize{TH Vs VH} & 1.790 & 3.346 & 1.557 & 310 & 310 & -8.103 & 0.000 \\
PDCCore & Texture-based & \footnotesize{TH Vs VP}  & 2.329 & 2.748 & 0.420 & 243 & 1066 & -8.663 & 0.000 \\
 & Texture-based & \footnotesize{TH Vs VH} & 2.329 & 2.672 & 0.343 & 243 & 81 & -6.109 & 0.000 \\
 & Contrast-based & \footnotesize{TH Vs VP}  & 3.183 & 8.089 & 4.906 & 310 & 157 & -7.428 & 0.000 \\
 & Contrast-based & \footnotesize{TH Vs VH} & 3.183 & 4.648 & 1.465 & 310 & 310 & -2.426 & 0.016 \\
\bottomrule
\end{tabular}

		\begin{tablenotes}
			\item{TH = Train healthy, VH = Test/validate healthy, VP = Test/validate pathological}
			\item{Subscript 1 refers to TH distribution and subscript 2 is the other distribution in the test}
			\item{Largest absolute mean difference is highlighted in bold font}
		\end{tablenotes}
	\end{threeparttable}
\end{table}

\emph{A reasonably capable underlying autoencoder with in-batch population-level context.} 
The baseline autoencoder, PDCCore, seems capable of estimating sufficient levels of detail and is not collapsing to an identity mapping of its inputs. Moreover, scoping the population context within the training mini-batches appears feasible as the PopuSense module does not seem to break the core functionality of the PDC, which is additionally inferred from the ~\CGLOWPOP~ configuration having similar performance to the baseline. 

\emph{Does the PopuSense approach improve model performance?} 
Applying \CGHIGHPOP~to \CONTRASTBASED-based input attains the best separability between healthy and pathological distributions, demonstrating the utility of the proposed method in enhancing the representations learned by a pathology detection convolutional autoencoder. Performance under \TEXTUREBASED-based input suggests that the vanilla version of \ADHGMODULE, a version without extra guidance or optimization support, seems inadequate for textural input. Overall, \ADHGMODULE~addresses the separability problem for contrast-based images with minimal adaptation, and requires further investigation for textural input and deterring reconstruction of anomalous regions.

\emph{How extensive should the reach of the population context be; large or small?} 
A wide-reaching population-level context yields better performance than a narrow one. It could be that the smaller context lacks adequate input to learn additional intra-group properties and thus is no different from the baseline model.

Applying \CGHIGHPOP~to \CONTRASTBASED-based input attains the best separability between healthy and pathological distributions. Moreover, a wide-reaching population-level context yields better performance than a small one. This demonstrates the utility of the proposed method in enhancing the representations learned by a pathology detection convolutional autoencoder. Furthermore, performance under \TEXTUREBASED-based input suggests that the vanilla version of \ADHGMODULE, a version without extra guidance or optimization support, seems inadequate for textural input. Overall, \ADHGMODULE~addresses the separability problem for contrast-based images with minimal adaptation, and requires further investigation for textural input

\textbf{Limitations.} 
The presented results demonstrate a proof of concept and may benefit from additional interrogation with a variety of input signals and contextualized hyperparameters and optimization functions. Furthermore, while we utilize datasets that constitute different data sources for diversity, additional considerations may be necessary when generalizing to significantly different imaging modalities due to the inherent selection bias in medical imaging datasets, where acquisition is not typical for healthy individuals.

\section{Conclusion}
\label{sec:adpconc}

This study presents a population-level context for pathology detection and takes on a graph theoretic approach to model and encode it into the latent representation of the underlying convolutional autoencoder. The aim is to enhance the code's capacity to capture additional intra-group variations inherent in biomedical data, which a local or global context of the convolutional model might miss or smooth out. Experimental results reveal improved separability between healthy and pathological observations in contrast-based medical images and struggle with extending that performance to textural input, which is a recurring challenge in existing research. All in all, such an approach avails an alternate avenue for enhancing the representations learned by a pathology detection convolutional model.


\section*{Acknowledgements} 
We thank the support from the National Natural Science Foundation of China 31970752; 32350410397; Science, Technology, Innovation Commission of Shenzhen Municipality JSGG20200225150707332, JCYJ20220530143014032, WDZC20200820173710001, WDZC20200821150704001; Shenzhen Medical Academy of Research and Translation, D2301002; Shenzhen Bay Laboratory Open Funding, SZBL2020090501004; Department of Chemical Engineering-iBHE special cooperation joint fund project, DCE-iBHE-2022-3; Tsinghua Shenzhen International Graduate School Cross-disciplinary Research and Innovation Fund Research Plan, JC2022009; and Bureau of Planning, Land and Resources of Shenzhen Municipality (2022) 207.

{
	\bibliographystyle{unsrtnat} 
	\bibliography{../tf-zadpaper}
}

\end{document}